\documentclass[fleqn,usenatbib]{mnras}

\usepackage[T1]{fontenc}

\DeclareRobustCommand{\VAN}[3]{#2}
\let\VANthebibliography\thebibliography
\def\thebibliography{\DeclareRobustCommand{\VAN}[3]{##3}\VANthebibliography}

\providecommand{\noopsort}[1]{}

\usepackage{graphicx}	
\usepackage{amsmath}	
\usepackage{amssymb}	

\usepackage{bm}

\usepackage{tikz,xcolor,hyperref}
\usepackage{graphicx}
\graphicspath{{images/}}

\definecolor{lime}{HTML}{A6CE39}
\DeclareRobustCommand{\orcidicon}{%
	\begin{tikzpicture}
	\draw[lime, fill=lime] (0,0) 
	circle [radius=0.16] 
	node[white] {{\fontfamily{qag}\selectfont \tiny ID}};
	\draw[white, fill=white] (-0.0625,0.095) 
	circle [radius=0.007];
	\end{tikzpicture}
	\hspace{-2mm}
}

\newcommand{\orcidVP}{\href{https://orcid.org/0000-0002-3031-062X}{\orcidicon}}
\newcommand{\orcidEV}{\href{https://orcid.org/0000-0003-2742-6872}{\orcidicon}}

\defcitealias{pav_ves_letter}{I}
\defcitealias{pav_ves_2}{II}

\newcommand{\Msun}{M_\odot}

\newcommand{\trh}[1][]{t_\mathrm{rh#1}}
\newcommand{\tcc}[1][]{t_\mathrm{cc#1}}
\newcommand{\ra}[1][]{r_\mathrm{a#1}}
\newcommand{\rh}[1][]{r_\mathrm{h#1}}
\newcommand{\rt}[1][]{r_\mathrm{t#1}}

\newcommand{\disp}[1][]{\sigma_\mathrm{#1}}

\newcommand{\Nb}[1][]{N_\mathrm{b#1}}
\newcommand{\Eb}[1][]{E_\mathrm{b#1}}
\newcommand{\n}[1][]{n_\mathrm{#1}}
\newcommand{\rb}[1][]{r_\mathrm{b#1}}

\newcommand{\der}[1]{\mathop{}\!\mathrm{d}#1}
\newcommand{\mm}[2]{\langle m \rangle^{#1\,\%}_{#2\,\%}}
\newcommand{\Vth}{V_\mathrm{th}}

\usepackage{newtxtext,newtxmath}

\title[Kinematics, mass segregation and binaries]{Mass segregation and dynamics of primordial binaries in star clusters with a radially anisotropic velocity distribution}

\author[Pavl\'ik \& Vesperini]{
V\'aclav Pavl\'ik\,$^{1}$\thanks{E-mail: vpavlik@iu.edu}\orcidVP,
Enrico Vesperini\,$^{1}$\orcidEV
\\
$^{1}$Indiana University, Department of Astronomy, Swain Hall West, 727 E 3$^\text{rd}$ Street, Bloomington, IN, 47405, USA
}

\date{Accepted 2022 June 23. Received 2022 June 21; in original form 2022 May 04}

\pubyear{2022}

\begin{document}
\label{firstpage}
\pagerange{\pageref{firstpage}--\pageref{lastpage}}
\maketitle

\begin{abstract}
	This paper is the third in a series investigating, by means of $N$-body simulations, the implications of an initial radially anisotropic velocity distribution on the dynamics of star clusters.
	Such a velocity distribution may be imprinted during a cluster's early evolutionary stages and several observational studies have found examples of old globular clusters in which radial anisotropy is still present in the current velocity distribution.
	Here we focus on its influence on mass segregation and the dynamics of primordial binary stars (disruptions, ejections, and component exchanges).
	The larger fraction of stars on radial/highly eccentric orbits in the outer regions of anisotropic clusters lead to an enhancement in the dynamical interactions between inner and outer stars that affects both  the process of mass segregation and the evolution of primordial binaries.
	The results of our simulations show that the time scale of mass segregation of the initially anisotropic cluster is longer in the core and shorter in the outer regions, when compared to the initially isotropic system.
	The evolution of primordial binaries is also significantly affected by the initial velocity distribution and we find that the rate of disruptions, ejections and exchange events affecting the primordial binaries in the anisotropic clusters is higher than in the isotropic ones.
\end{abstract}

\begin{keywords}
globular clusters: general -- stars: binaries: general -- stars: kinematics and dynamics -- methods: numerical
\end{keywords}


\newlength{\colfigwidth}
\setlength{\colfigwidth}{\linewidth}


\section{Introduction}
\label{sec:intro}

Many recent observational studies have provided new insights into the internal kinematics of star clusters \citep[SCs; see e.g.][]{bellini_hstV,libralato_hstIII,libralato_hst,ferraro_etal,MUSE,jindal_gaia,2022arXiv220605300W} and led to a significant progress towards building their complete dynamical picture.
They show that SCs are often characterised by velocity anisotropy or internal rotation, at odds with the traditionally adopted description of these systems based on isotropic and non-rotating models.
These kinematic properties are likely to emerge during the formation and early evolutionary phases \citep[e.g.][]{vesperini_etal14,mapelli17,lahen20,livernois21,Tio_Ves_Var22} and are subsequently modified during the SCs long-term evolution by the effects of two-body relaxation and the external tidal field of the host galaxy \citep[e.g.][]{Tio_Ves_Var17,Tio_Ves_Var18,Tio_Ves_Var22}.
It has also been reported that rotation \citep[e.g.][]{fokkerplanck_rotI,fokkerplanck_rotII,fokkerplanck_rotIII,hong_etal,Tio_Ves_Var17,Tio_Ves_Var19,livernois22} and anisotropic velocity distributions \citep[e.g.][]{breen_var_heg,pav_ves_letter,pav_ves_2} have significant impact on the evolution of the structural and dynamical properties of SCs.

In the first two papers of the series \citep[][hereafter Papers~\citetalias{pav_ves_letter} \&~\citetalias{pav_ves_2}]{pav_ves_letter,pav_ves_2}, we have explored the role of stellar kinematics on the SCs evolution towards energy equipartition (including its dependence on the strength of the external tidal field and the fraction of primordial binary stars). Our results showed significant differences between the SCs with isotropic and radially anisotropic initial velocity distributions.
In this paper we extend our work to further investigate the effects of a radially anisotropic velocity distribution on two fundamental processes in the long-term dynamical evolution of SCs: mass segregation and the dynamics of primordial binary stars.

Mass segregation is one of the main manifestations of two-body relaxation and significantly affects the dynamics of self-gravitating systems composed of stars with unequal masses \citep[e.g.][]{spitzer,heggie_hut}. It influences the variation of the global and local stellar mass function \citep[see e.g.][]{ves_heggie97,baum_makino,gill_etal}, and the evolution of the cluster's stellar content, including the formation of binary stars while its central region undergoes core collapse \citep[e.g.][]{aarseth1972,hut,fujii_pz,oleary,pavl_subr}. Primordial binaries play a key role in the structural development of a SC since they act as heating sources halting core collapse earlier and at lower densities \citep[see e.g.][]{heggie_tren_hut,tre_heg_hut,bin_stable,ves_cher94}. Moreover, stellar encounters involving binary stars may result in the formation of a variety of exotic objects such as blue stragglers, binary black holes, X-ray binaries, millisecond pulsars etc., which could also have impact on the Galactic scale \citep[see e.g.][and references therein]{ivanova_etal06,shara_hurley,hypki,macleod,hong_etal17,kremer18,hong_etal18,Varri2018,ye20,2020PASA...37...44A,weatherford21,msp,fra22}.

The results presented in this paper show that both mass segregation and the dynamics of binary stars encounters are significantly affected by the SC's initial kinematic properties. The outline of the paper is as follows: in Section~\ref{sec:methods} we describe our models and methods, in Section~\ref{sec:segr} we present the results concerning mass segregation and in Section~\ref{sec:bins} those describing the evolution of primordial binary stars, and our conclusions are summarised in Section~\ref{sec:concl}.

\section{Methods}
\label{sec:methods}

\begin{figure}
	\centering
	\includegraphics[width=\linewidth]{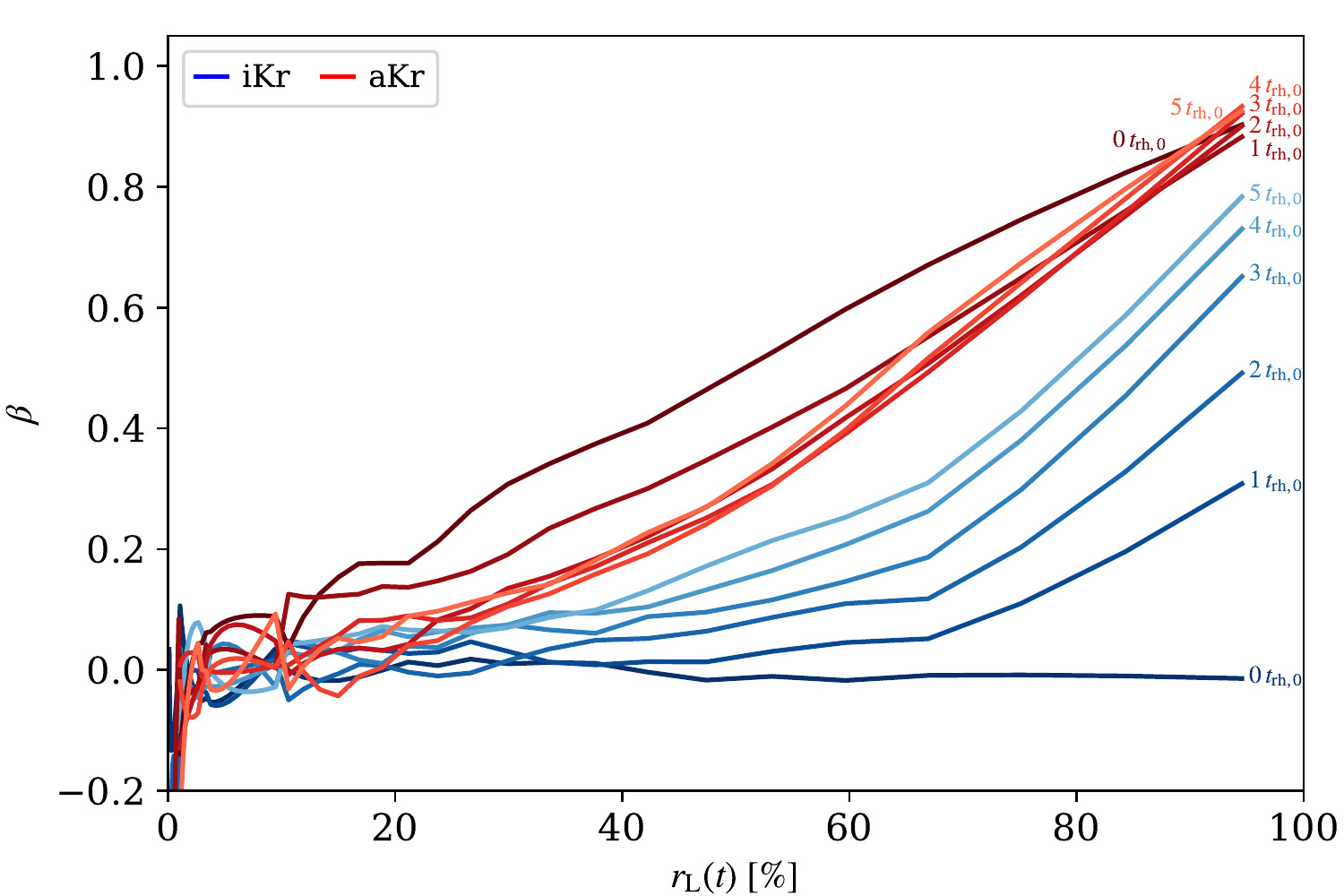}
	\caption{Radial dependence of the velocity anisotropy in the modelled \texttt{Kr} clusters (colour-coded) measured with the parameter $\beta$, see Eq.~\eqref{eq:aniso}. The line shading corresponds to the evolutionary time in $\trh[,0]$, see Eq.~\eqref{eq:trh}, and the radial position is given by the Lagrangian radii of a given mass percentage at each time.}
	\label{fig:beta}
\end{figure}

We studied star clusters with $N_0 = 1.1 \times 10^5$ stars, represented by $N$-body models with an initial \citet{king_model} density profile (using the central dimensionless potential $W_0 = 6$).
The modelled clusters were placed in a weak point-like external potential with a ratio of the tidal radius to the initial cluster truncation radius, $\rt / r_\mathrm{cl} = 10$ (we note that the initial tidal radius is ${\approx}70$~half-mass radii). The models were evolved numerically with the direct $N$-body integrator \textsc{nbody6++gpu} \citep{nbody6pp}.

As in Papers~\citetalias{pav_ves_letter} and \citetalias{pav_ves_2}, we considered two distinct initial velocity distributions with a radial variation of the anisotropy following the Osipkov--Merritt profile \citep{osipkov,merritt}
\begin{equation}
	\label{eq:aniso}
	\beta = 1 - \frac{\disp[tan]^2}{2 \disp[rad]^2} = \frac{(r/\ra)^2}{1 + (r/\ra)^2} \,,
\end{equation}
where $\disp[rad]$ and $\disp[tan]$ are the radial and tangential components of the velocity dispersion, respectively \citep[implemented numerically with \textsc{agama}][]{agama}.
In the regions up to the anisotropy radius, $\ra$, all clusters have an approximately isotropic velocity distribution, and they are radially anisotropic beyond $\ra$.
We compared the evolution of isotropic models which, by definition, have \hbox{$\ra{\rightarrow}\infty$} (hereafter, we refer to these models using labels beginning with~`\texttt{i}'), and radially anisotropic models. For those, we adopted $\ra = \rh[,0]$\,, where $\rh[,0]$ is the initial half-mass radius (their labels start with~`\texttt{a}').
Such a radial variation of anisotropy in the velocity distribution is characteristic for stellar systems at the end of the early evolution and violent relaxation phases \citep[see e.g.][]{vanAlbada,tre_ber_van,vesperini_etal14}. In Fig.~\ref{fig:beta} we show the initial radial profiles of $\beta$ for the two distributions considered in this study, and their evolution at a few representative times. The initially isotropic model gradually develops a radially anisotropic velocity distribution in the outer regions as a result of the outward diffusion of stars driven by the effects of two-body relaxation. The initially anisotropic model, on the other hand, shows only a moderate evolution of $\beta$. This is consistent with the findings of \citet{Tio_Ves_Var16} and we refer to that paper for a study specifically aimed at exploring the long-term evolution of the anisotropy for systems with a variety of different initial structural and kinematical properties.

Since we primarily focus on the dynamical effects associated with binary encounters, two-body relaxation, and the role of velocity anisotropy in these processes, our simulations do not include stellar evolution and start from a more idealised initial setup. In all models, stellar masses were drawn from a narrow range which is typical of an old star cluster, specifically using the \citet{kroupa} initial mass function (IMF) between $0.1 \leq m / \Msun \leq 1.0$ (denoted by~`\texttt{Kr}', in the labels).

To further investigate some aspects of the evolution of primordial binary stars in Section \ref{sec:bins} (in particular the rate of encounters resulting in the exchange of one binary component with a high-mass star), we also used models with only two distinct mass components (labelled `\texttt{2c}').
To approximate the number and masses of the high-mass stars -- which would be represented by black holes (BHs) and neutron stars (NSs) -- we evolved the \citet{kroupa} IMF with $N_0$ stars with masses between $0.08$ and $150\,\Msun$ for 12\,Gyr. We used the parametric single-stellar evolutionary algorithms \citep{sse,bse_bel2002} implemented in \textsc{Mcluster} \citep[][]{mcluster}, assuming the average metallicity of globular clusters in the Galaxy, $\left[ \mathrm{Fe}/\mathrm{H} \right] = -1.30$ \citep[e.g.][]{baumgardt_sollima}. We ended up with 830 dark stellar remnants with a mean mass ${\approx}13.4$ times higher than the mean mass of the remaining stars. Therefore, our two-component models are composed of two groups of stars with a mass ratio $13.4:1$, with 830 massive stars, and the remainder up to $N_0$ are equal-mass low-mass stars.
We note that in this simple set up, we did not consider supernova kicks or any dynamical effects (such as mass loss due to ejections or evaporation of stars) that would affect the actual fraction of dark remnants at 12 Gyr.
Although the population of dark remnants may, therefore, be overestimated when compared to some real system, our idealised setup is sufficient for the purpose of comparing the relative numbers of stellar interactions in the isotropic and anisotropic clusters.

In all models, $2\,\Nb[,0]$ out of the $N_0$ point-mass stars were paired into primordial binaries. Specifically, the total number of binary systems was set to $\Nb[,0]=10^4$.
The binary population had a uniform distribution in the log of the binding energy between $0.1$ and $30\,kT$ and a thermal distribution of eccentricities $f(e)=2e$ \citep[see e.g.][]{heggie75}.
In the \texttt{Kr} models, the secondary masses were drawn from a uniform distribution between $0.1\,\Msun$ and the primary mass; in the \texttt{2c} models, only the corresponding fraction of the low-mass components was put in equal-mass binaries.

The time development of the models is expressed with two dynamical timescales. Either in units of the initial half-mass relaxation time,
\begin{equation}
	\label{eq:trh}
	\trh[,0] = \frac{0.138 \, N_0\,\rh[,0]^{3/2}}{\ln{(0.02 N_0)}} \,,
\end{equation}
\citep[all in H\'enon units, see e.g.][]{spitzer}, or in units of the time of core collapse, $\tcc$. While $\trh[,0]$ is the same for all models, $\tcc$ depends on the evolution of the particular model and can be estimated by different methods \citep[e.g.][]{makino,pavl_subr} or visually identified close to the minimum of a low-percentage Lagrangian radius (e.g.~$1\,\%$, see Fig.~\ref{fig:rl_Kr}).

\section{Results: Mass segregation}
\label{sec:segr}

\begin{figure}

	\centering
	\includegraphics[width=\linewidth]{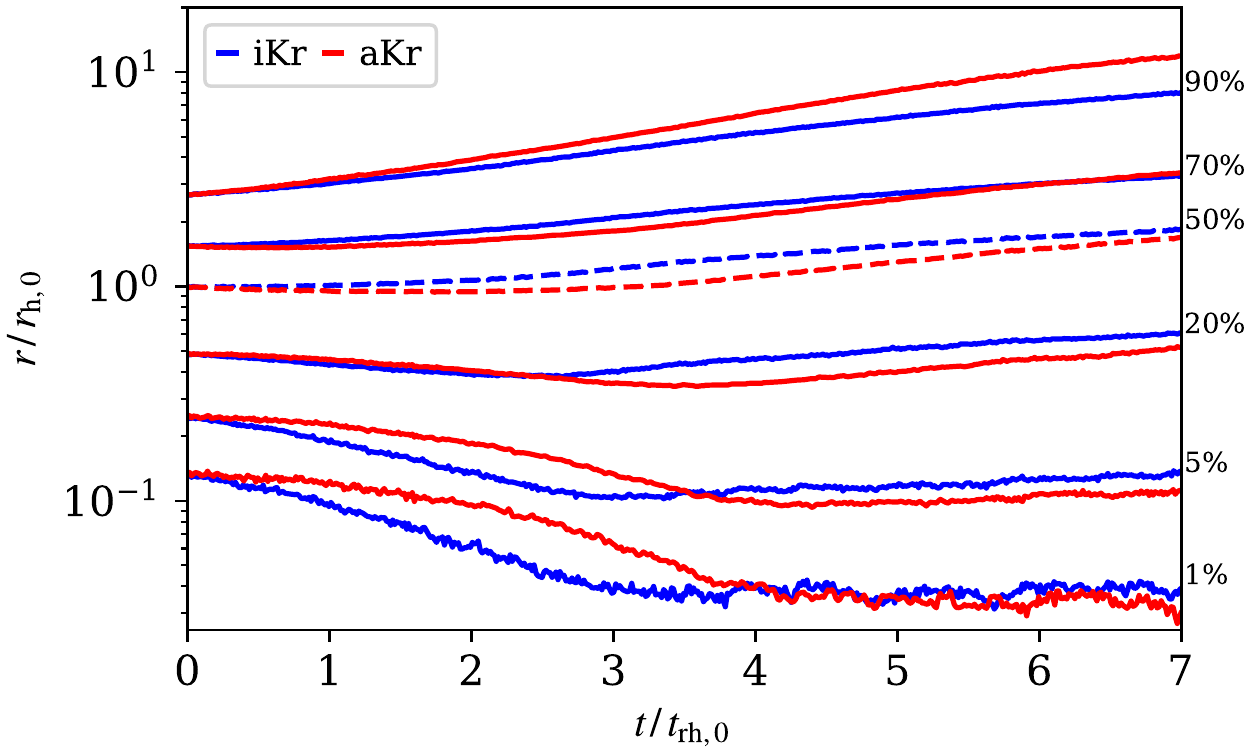}
	\caption{Time evolution of our isotropic (blue) and anisotropic (red) models with the Kroupa IMF shown by the 1, 5, 20, 50 and 70\,\% Lagrangian radii (counted from all stars up to the tidal radius and normalised to the initial half-mass radius). The lines show the raw data, with snapshots generated every $0.007\,\trh[,0]$.}
	\label{fig:rl_Kr}
	
	\vspace{\floatsep}
	
	\includegraphics[width=\linewidth]{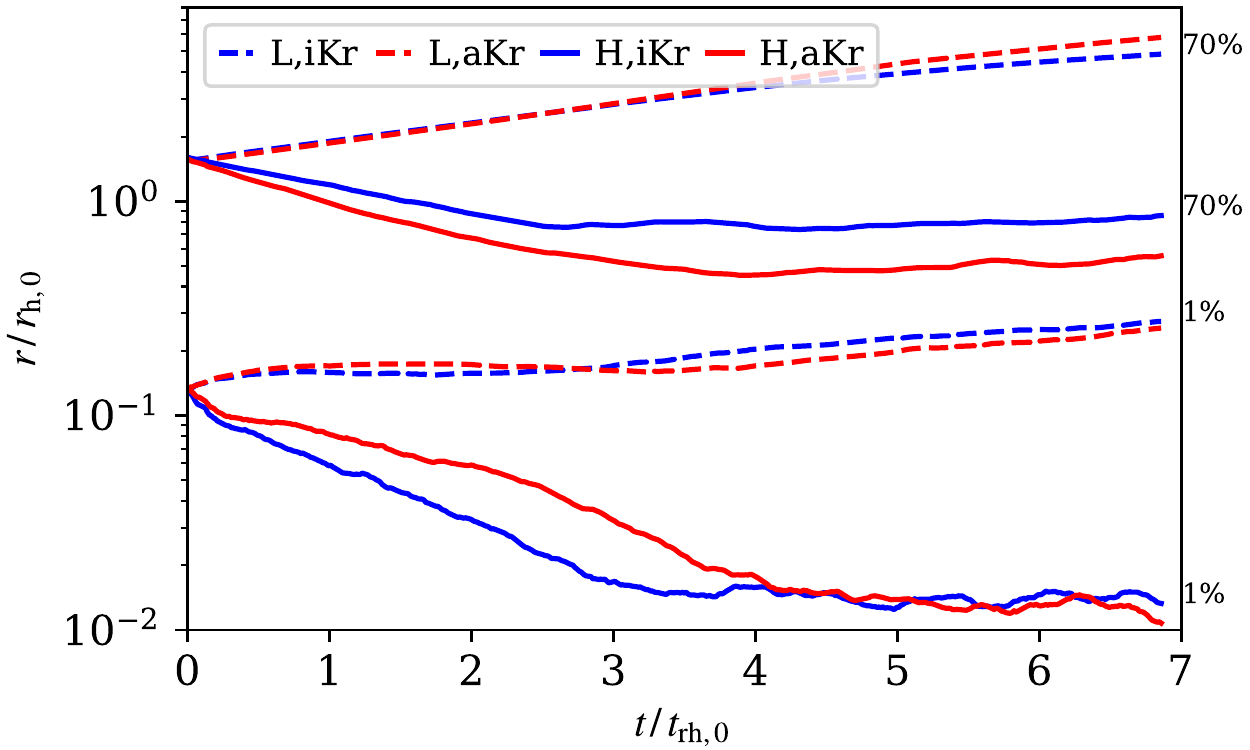}
	\caption{The inner (1\,\%) and outer (70\,\%) Lagrangian radii of the isotropic and the anisotropic models with the Kroupa IMF. Low-mass (dashed) and high-mass stars (solid) are shown separately. The lines have been smoothed by a moving average of window length $0.148\,\trh[,0]$ to better illustrate the global trends.}
	\label{fig:rl_Kr_LH}
	
	\vspace{\floatsep}
	
	\includegraphics[width=\linewidth]{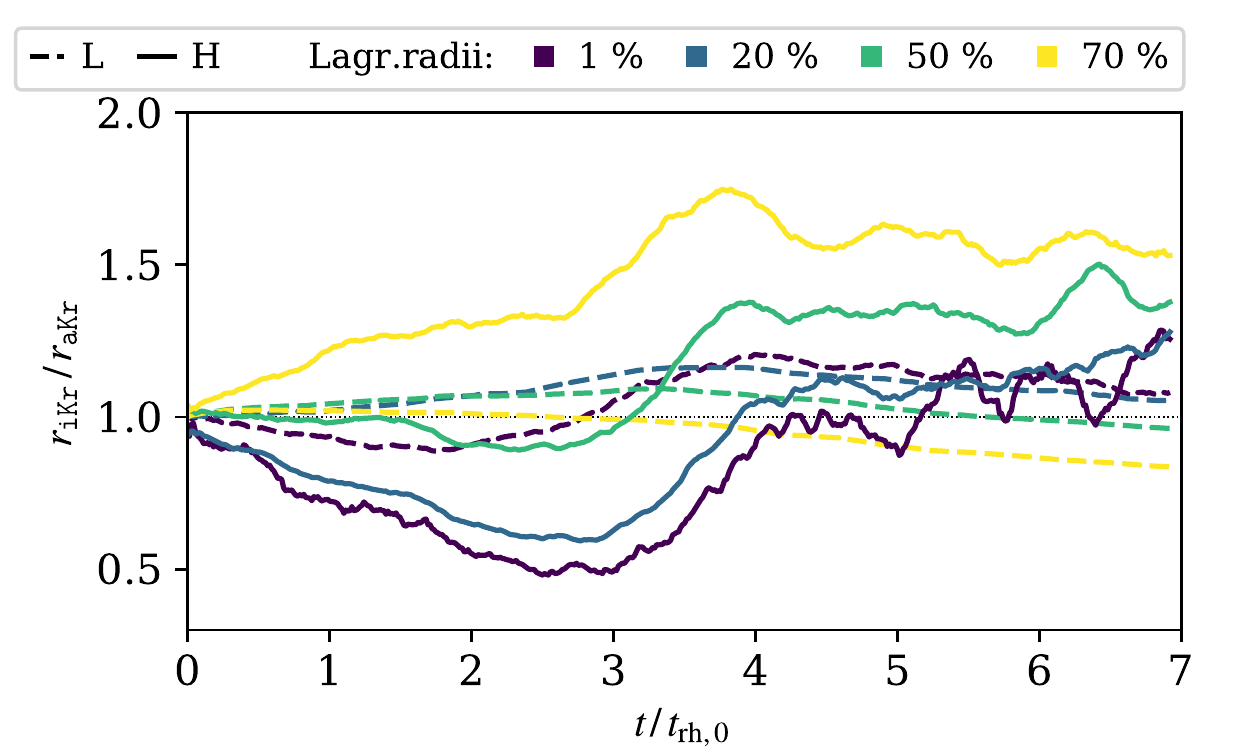}
	\caption{Ratio of the Lagrangian radii of the isotropic and the anisotropic model with the Kroupa IMF. Low and high mass stars are evaluated separately (see text) and the mass percentage is distinguished by colour. The lines have been smoothed by a moving average of window length $0.148\,\trh[,0]$ to better illustrate the global trends.}
	\label{fig:rl_Kr_LLHHratio}
\end{figure}

\begin{figure}
	\centering
	\includegraphics[width=\linewidth]{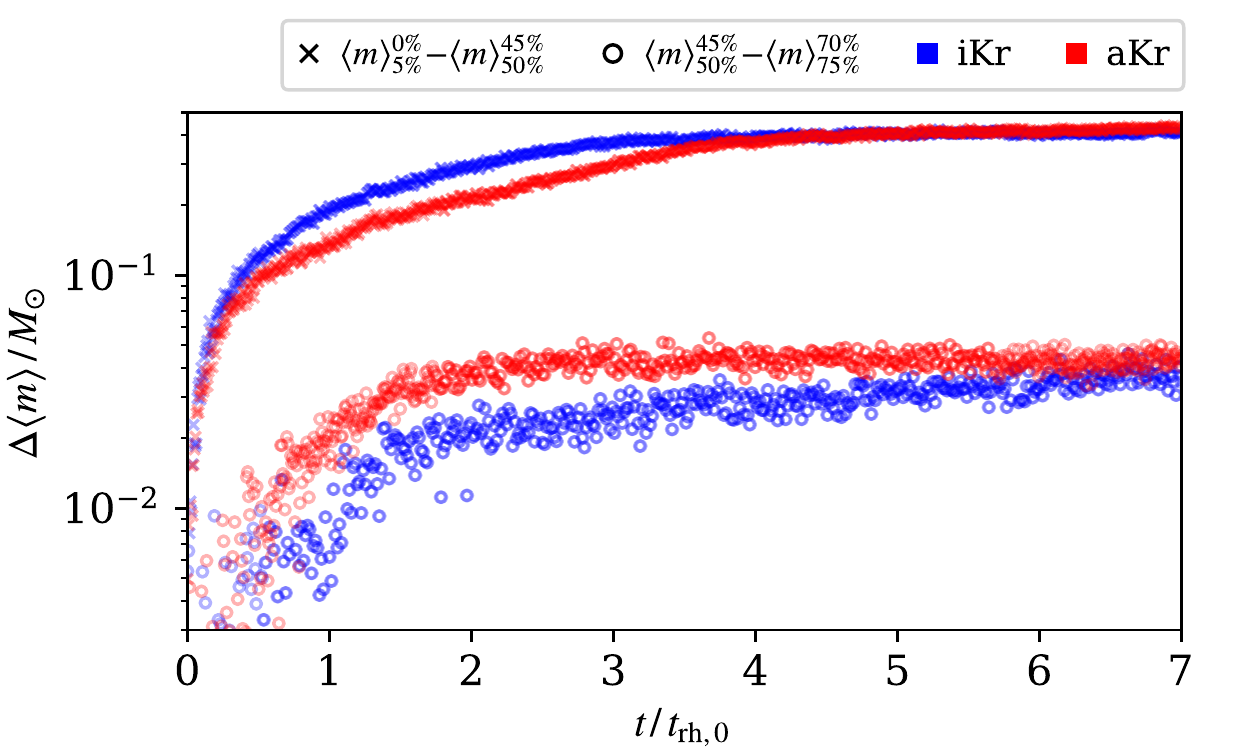}
	\caption{Evolution of mass segregation in our models (colour-coded). Separate comparisons between the core and the half-mass radius (crosses), and between the half-mass radius and the halo (circles) are plotted as a difference of the mean stellar mass comprised in these regions. The ranges of Lagrangian radii from which the mean mass is taken are specified as mass percentages in the legend.}
	\label{fig:deltam}
\end{figure}

Here we focus on the process of mass segregation and explore how it is affected by the initial radial velocity anisotropy.
First, we separately evaluate the spatial distribution of the high-mass stars ($0.9 \leq m_\mathrm{H}/\Msun \leq 1.0$, labelled ``H'') and the low-mass stars ($0.1 \leq m_\mathrm{L}/\Msun \leq 0.2$, labelled ``L''). We show the inner and outer Lagrangian radii of both mass groups in Fig.~\ref{fig:rl_Kr_LH}, and the ratios of various Lagrangian radii for the \texttt{iKr} and \texttt{aKr} models in Fig.~\ref{fig:rl_Kr_LLHHratio}.
The overall evolution of the low-mass stars is almost the same in both models (see the overlapping dashed lines in Fig.~\ref{fig:rl_Kr_LH}; and the dashed lines in Fig.~\ref{fig:rl_Kr_LLHHratio} staying very close to $r_\texttt{iKr} \big/ r_\texttt{aKr} = 1$). However, both figures clearly show a significant difference between the segregation of massive stars: it proceeds more rapidly in the inner regions of the isotropic cluster, and the opposite behaviour is found in the outer regions where the inward migration of massive stars is more rapid for the anisotropic system.

The evolution of the 1 per cent Lagrangian radii in Figs.~\ref{fig:rl_Kr} \&~\ref{fig:rl_Kr_LH} also displays the difference between the time necessary to reach core collapse in the two models (see also \citealt{breen_var_heg} and Papers~\citetalias{pav_ves_letter} \&~\citetalias{pav_ves_2} for the previous studies on the dependence of the core collapse time on the radial anisotropy in the velocity distribution).
Consequently, we can see a non-monotonic behaviour of the time evolution of the inner Lagrangian radii ratios in Fig.~\ref{fig:rl_Kr_LLHHratio}. Specifically, the ratio of the 1\% radii of the high-mass stars initially steadily decreases (and a similar trend is observable at least up to the radius containing the innermost 20\,\% mass of the high-mass stars) as a result of the massive component of the isotropic model segregating more rapidly in the inner regions. At about $3\,\trh[,0]$ (coinciding with the \texttt{iKr} going through core collapse and stopping its central contraction), this ratio reaches its minimum and then starts to increase again. The inner regions of the anisotropic model, on the other hand, are still slowly contracting. At approximately $4\,\trh[,0]$, when the \texttt{aKr} model also reaches core collapse, the cores of both SCs become similar and the ratios stabilise to values approximately equal to one.

We interpret this delayed mass segregation in the core of the anisotropic system compared to the isotropic one (and vice versa in the outer region) as the result of the differences in the orbit distribution of their constituent stars.
A larger fraction of the outer stars in the \texttt{aKr} model are characterised by eccentric/radial orbits and are, therefore, subject to an increased number of interactions as they venture in the inner regions. This enhanced interaction rate accelerates their spatial segregation (as observed in Figs.~\ref{fig:rl_Kr_LH} \&~\ref{fig:rl_Kr_LLHHratio}). At the same time, they are also acting as a heating source and are delaying core collapse and mass segregation in the inner regions. 

To corroborate our results and further illustrate the differences in the mass segregation process in the two systems, we analyse the mean stellar mass in various regions of the clusters \citep[see e.g.][where the same diagnostic was used to explore differences in the segregation of SCs with and without an intermediate-mass black hole]{gill_etal}. In Fig.~\ref{fig:deltam}, we plot the time evolution of the difference in the mean mass in two spherical shells. The case of $\Delta m = \mm{0}{5} - \mm{45}{50}$ (i.e.\ between the a shell around the half-mass radius and the inner region, up to the 5\,\% Lagrangian radius) illustrates the spatial redistribution of stellar masses in the cluster core. The two sequences corresponding to the \texttt{iKr} and \texttt{aKr} models show a more rapid development of mass segregation in the isotropic system, persisting until ${\approx}4\,\trh[,0]$. This confirms what we reported above.
In turn, the points representing $\Delta m = \mm{45}{50} - \mm{70}{75}$ (i.e.\ the difference between the mean mass in an outer shell and the half-mass radius shell) confirm that mass segregation proceeds more rapidly in the halo of the anisotropic system when compared to the isotropic one.

\section{Results: Binary stars}
\label{sec:bins}

In this section we focus on another important aspect of collisional dynamics of star clusters and explore how the evolution and survival of binary stars is affected by the initial velocity distribution and the underlying orbit distribution.

\subsection{Semi-analytical estimate}

Before discussing the results on the evolution of binary stars in our $N$-body simulations, we present a semi-analytical calculation illustrating the expected impact of the differences in the degree of anisotropy in the initial velocity distribution on the evolution of binary stars. Specifically, we focus on binary disruption.
We use the ionisation rate obtained by \citet{hut_bahcall} with an analytical fit to the simulations of binary disruptions in binary--single encounters
\begin{equation}
	\label{eq:ion}
	\frac{1}{\n[b]} \frac{\der{\n[b]}}{\der{t}} = \n[s] \pi a^2 \Vth R(x) \,,
\end{equation}
with the function
\begin{equation}
	R(x) = \frac{32}{27} \sqrt{\frac{6}{\pi}} \ \left[ \Big(1 + \frac{0.2 A}{x} \Big) \Big(1 + \exp{\frac{x}{A}} \Big) \right]^{-1} .
\end{equation}
The binary semi-major axis is denoted with $a$, the binary population has the number density $\n[b]$, and $\n[s]$ is the number density of the population of single stars interacting with the binaries. The hardness factor is $x = |\Eb| \big/ (m \sigma^2)$, where $E_b$ is the binary binding energy, $m$ is the mass of single stars, and $\sigma$ is their 1D velocity dispersion.  The dispersion of the binary--single stars relative velocities is
\begin{equation}
	\Vth = 3  \sigma \sqrt{A/2} \,,
\end{equation}
where $A = 1$ for the case of energy equipartition, and $A = 4/3$ for the case of velocity equipartition.

\begin{figure}
	\centering
	\includegraphics[width=\linewidth]{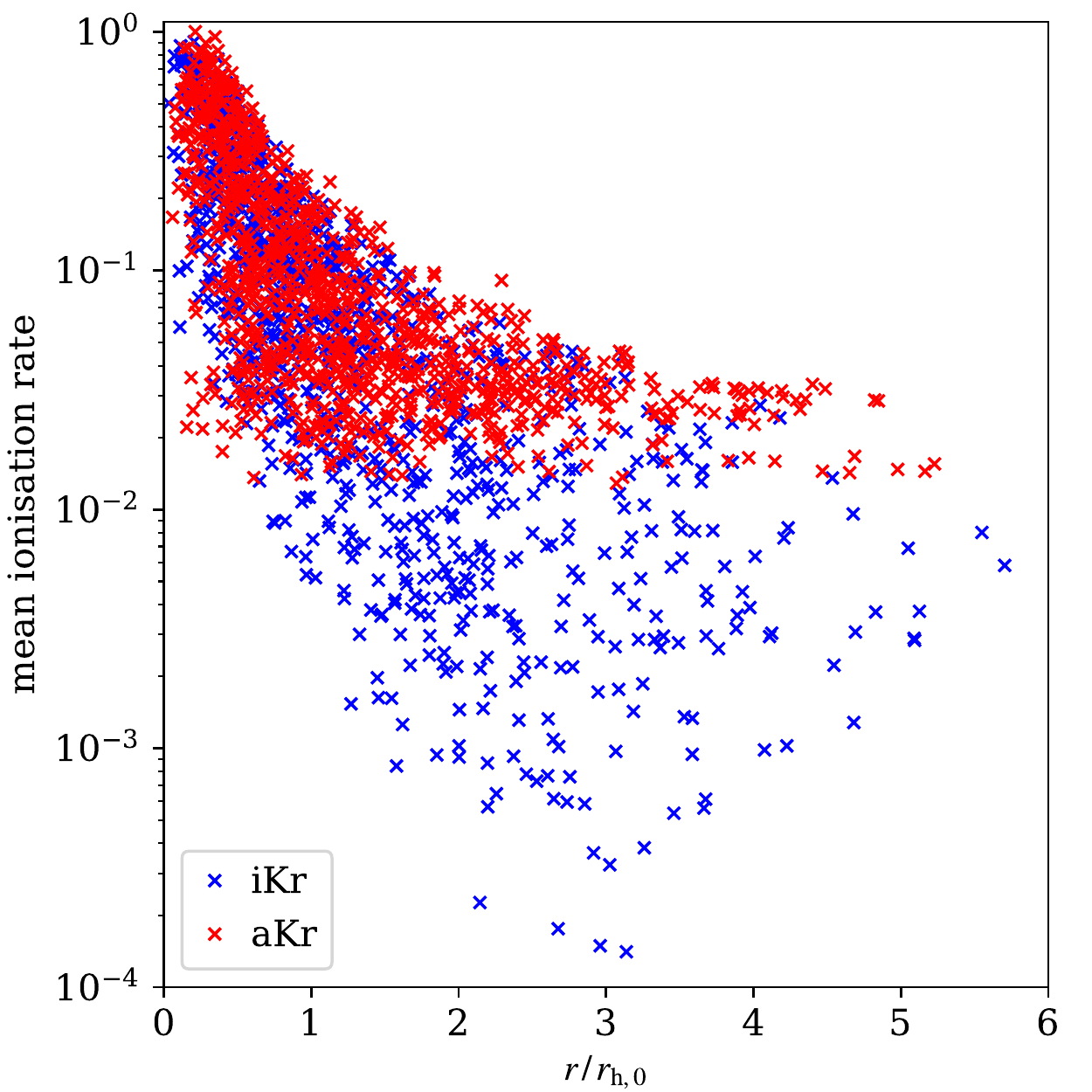}
	\caption{Mean ionisation rate (normalised to the maximum value in the set) in the isotropic (blue) and the anisotropic models (red) as a function of the initial radial positions of the stars in the cluster (in units of $\rh[,0]$). The plot is based on $10^3$ randomly selected stars in each model.}
	\label{fig:ion}
\end{figure}

We selected a random sample of $10^3$ stars in each of the \texttt{aKr} and \texttt{iKr} models and followed their orbits for 25 $N$-body time units (i.e.\ ${\approx}9$ crossing times which enabled even the outermost stars to complete at least one full orbit in the cluster). For each star we determined the local number density and velocity dispersion along its orbit, and used these quantities to calculate the mean ionisation rate from Eq.~\eqref{eq:ion}. For this numerical calculation we used the value of binding energy $|\Eb| = kT$, however, we note that the specific choice of $\Eb$ (or $a$) is not relevant for the general purposes of illustrating the differences in the ionisation rate between isotropic and anisotropic systems.

As we show in Fig.~\ref{fig:ion}, no significant differences are found between the ionisation rates of binary stars in the inner regions where both clusters are characterised by a similar isotropic velocity distribution. Outside the half-mass radius, however, the velocity distribution of the \texttt{aKr} model becomes increasingly radially anisotropic with a larger fraction of radial orbits plunging in the central regions where the local ionisation rate is higher.
Consequently, for $r \gtrsim \rh$ the two models differ: while the average ionisation rate for the stars in the isotropic system continues to decline, that of the anisotropic SC stays approximately constant.

Although this calculation only concerns the ionisation rate, the result we obtained is a general consequence of the enhanced interaction rate of binary stars on radial orbits.
Therefore, we expect to see similar differences for all processes affecting the evolution of the outer binary stars in the two models. In the next sections we present the analysis of our $N$-body simulations, concerning disruptions, ejections, and binary component exchanges.

\subsection{Evolution of the number of binary stars}

\begin{figure}
	\centering
	\includegraphics[width=\linewidth,trim=0 17 0 0,clip]{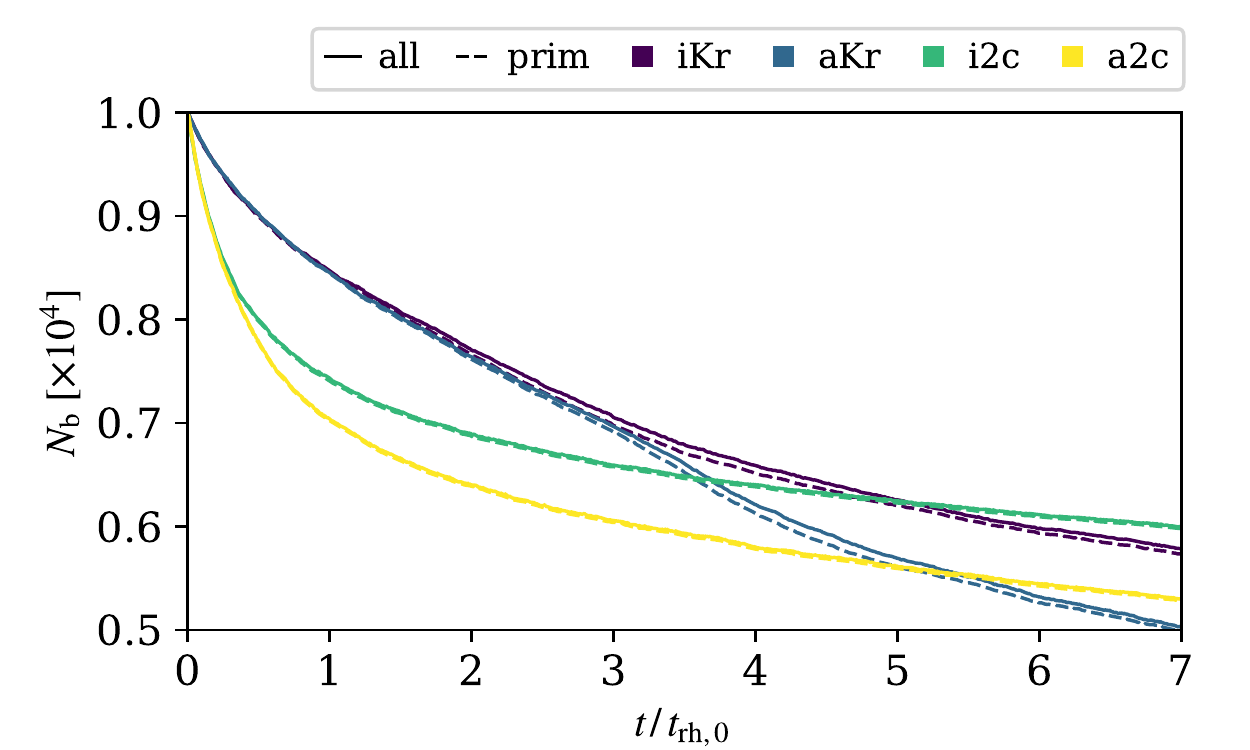}
	\includegraphics[width=\linewidth,trim=0 0 0 30,clip]{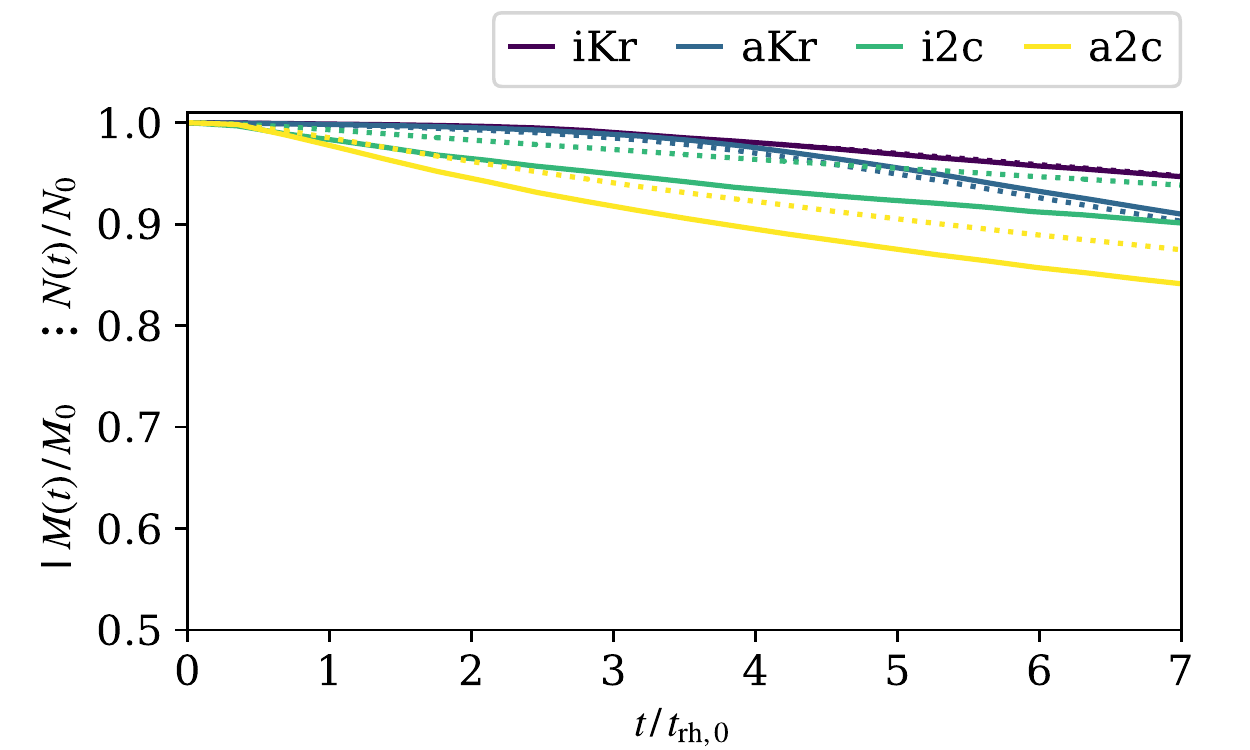}
	\caption{\textbf{Top:} Time evolution of the total number of binary stars (solid) and the number of surviving primordial binary stars (dashed) inside the tidal radius. The count of primordial binaries includes also the pairs where exchanges of components took place. \textbf{Bottom:} Time evolution of the total mass (solid) and the total number of stars (dotted) inside the tidal radius, normalised to the initial value.}
	\label{fig:Nb_Rb}
\end{figure}

\begin{figure*}
	\centering
	\includegraphics[width=\linewidth]{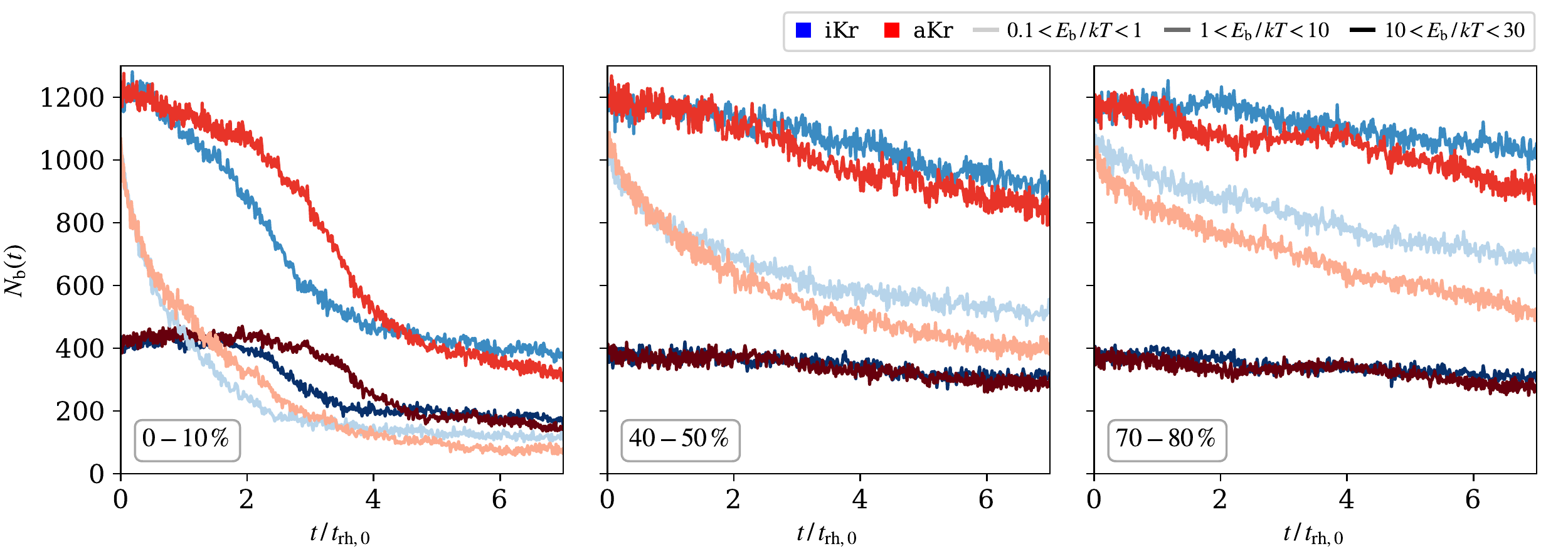}\\
	\caption{Evolution of the number of binaries in our models with an IMF (colour-coded) with a given binding energy (colour shade) in three radial shells (columns), defined by the Lagrangian radii of the specified mass percentage. Time is in units of the half-mass relaxation time, see e.g.\ Eq.~\eqref{eq:trh}.}
	\label{fig:Ebin_Trh}

	\vspace{\floatsep}
	
	\includegraphics[width=\linewidth]{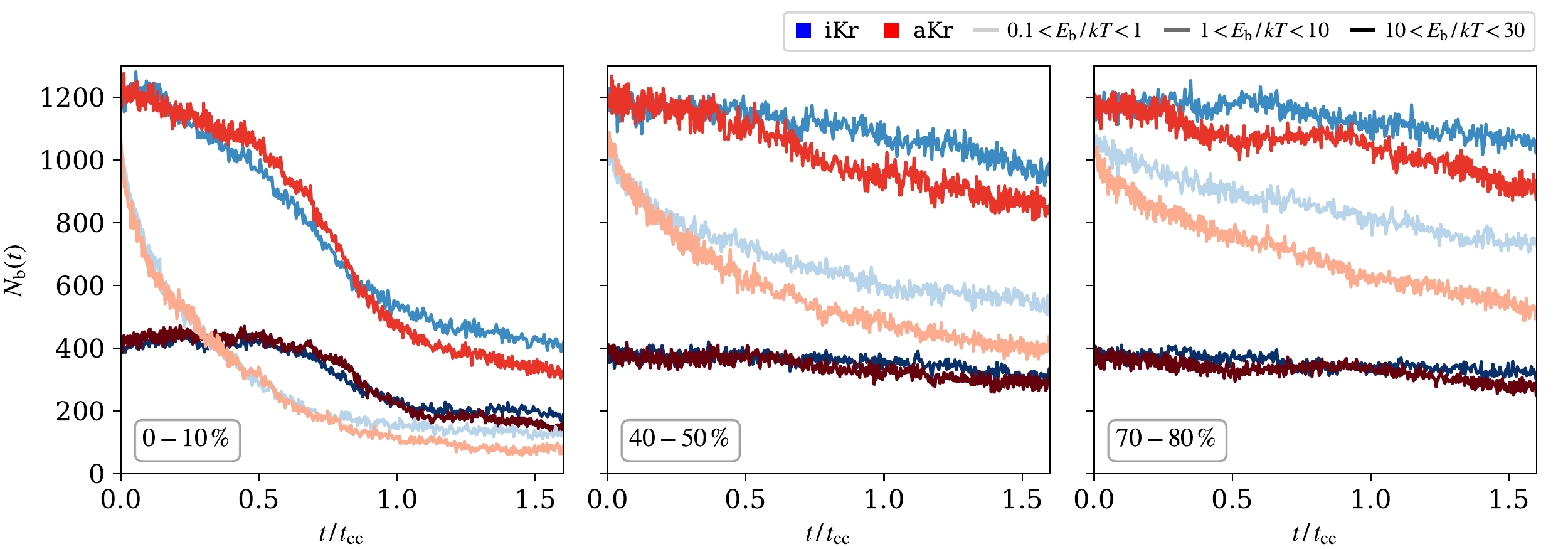}
	\caption{Same as Fig.~\ref{fig:Ebin_Trh} but time is in units of the times of core collapse of each model.}
	\label{fig:Ebin_Tcc}
\end{figure*}

The time evolution of the total number of binary stars and primordial binaries inside the tidal radius is shown for all the models in Fig.~\ref{fig:Nb_Rb}.
If a primordial binary exchanges one of its components after an encounter with another star (or binary), we are still counting it as part of the former primordial binary family. If, however, a primordial binary exchanges both components between two model outputs -- either as a result of a more complicated interaction of multiple stars, or two quick single exchanges -- it is no longer counted in the population of primordial binaries.
In the top panel of Fig.~\ref{fig:Nb_Rb}, we can see only a marginal difference between the number of primordial binaries and the total number of binaries at any given time. Hence, we conclude that these double exchanges or the formation of new binaries from random few-body interactions are rare and only add up to less that $1\,\%$ of the total binary population.

In all simulations, the number of binary stars decreases during the cluster's evolution (see the top panel of Fig.~\ref{fig:Nb_Rb}). This is a consequence of binary disruptions and ejections (resulting from the recoil velocity following binary--binary and binary--single encounters, mainly in the central regions). The fact that escaping binary stars are a negligible fraction of all binaries lost is further evident from the bottom panel of Fig.~\ref{fig:Nb_Rb}, where we plot the total mass and number of stars in the system. The rate of mass loss is much smaller than the rate of binary loss in both \texttt{iKr} and \texttt{aKr} models.
Furthermore, the mass loss is independent of the initial velocity anisotropy (see also Paper~\citetalias{pav_ves_2}), so it cannot explain the different numbers of binaries between the \texttt{iKr} and \texttt{aKr} models that we see the top panel of Fig.~\ref{fig:Nb_Rb}.

Fig.~\ref{fig:Ebin_Trh} provides further insights into the evolution of binary stars by showing the evolution of the total number of binaries based on their binding energies and positions in the cluster. We separately focus on three radial shells (the core, the intermediate regions around the half-mass radius, and the halo) and three energy bins.
Our results show that the evolution of the number of binary stars is similar in the central regions of both models (left panel of Fig.~\ref{fig:Ebin_Trh}). The only discrepancy in due to the different times of core collapse which is further evident from Fig.~\ref{fig:Ebin_Tcc} where time is normalised to $\tcc$ of each model and all the curves coincide (see the left panel). This is expected since binaries in the denser core region are subject to more frequent stellar interactions, in particular in the late evolutionary stages when the SC approaches core collapse (see e.g.\ the sudden drop in the number of binaries around $t=1\,\tcc$).

The evolution of the total number of the most compact binaries of both \texttt{iKr} and \texttt{aKr} models remains similar in the intermediate and outer regions as well (see the middle and the right-hand panels of Figs.~\ref{fig:Ebin_Trh} \&~\ref{fig:Ebin_Tcc}). However, we find a significant difference between the evolution of the number of wider binaries in these regions -- their count decreases more rapidly in the anisotropic model than in the isotropic one (normalising time to $\tcc$ even increases the separation of the two curves). 
This as another consequence of the different distributions of stellar orbits in the isotropic and anisotropic systems (see also Sect.~\ref{sec:segr}).
Binary stars from the outer regions that are on radial/highly eccentric orbits undergo frequent interactions with single and binary stars as they pass through the cluster's inner regions. This implies that binaries in the anisotropic system, where these orbits are initially more numerous will have an enhanced rate of disruption and ejection compared to the isotropic model.

\begin{figure}
	\centering
	\includegraphics[width=\linewidth,trim=0 17 0 0,clip]{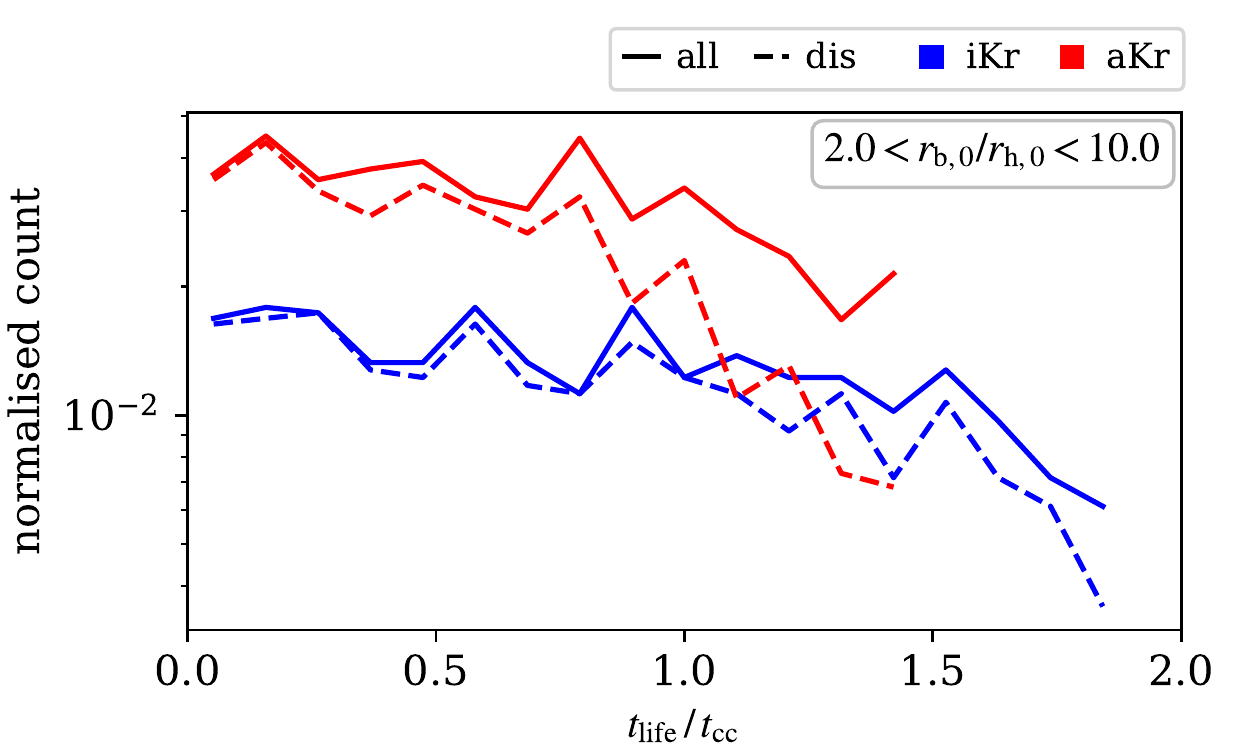}\\
	\includegraphics[width=\linewidth,trim=0 17 0 30,clip]{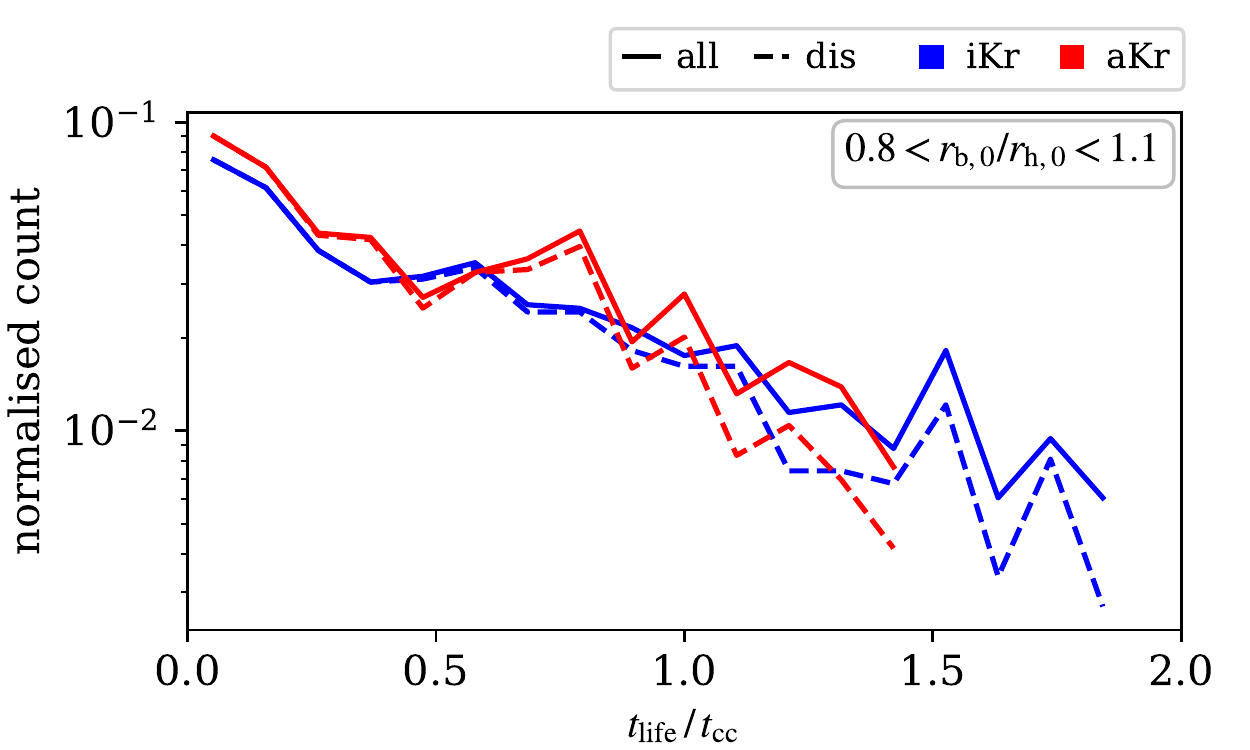}\\
	\includegraphics[width=\linewidth,trim=0 0 0 30,clip]{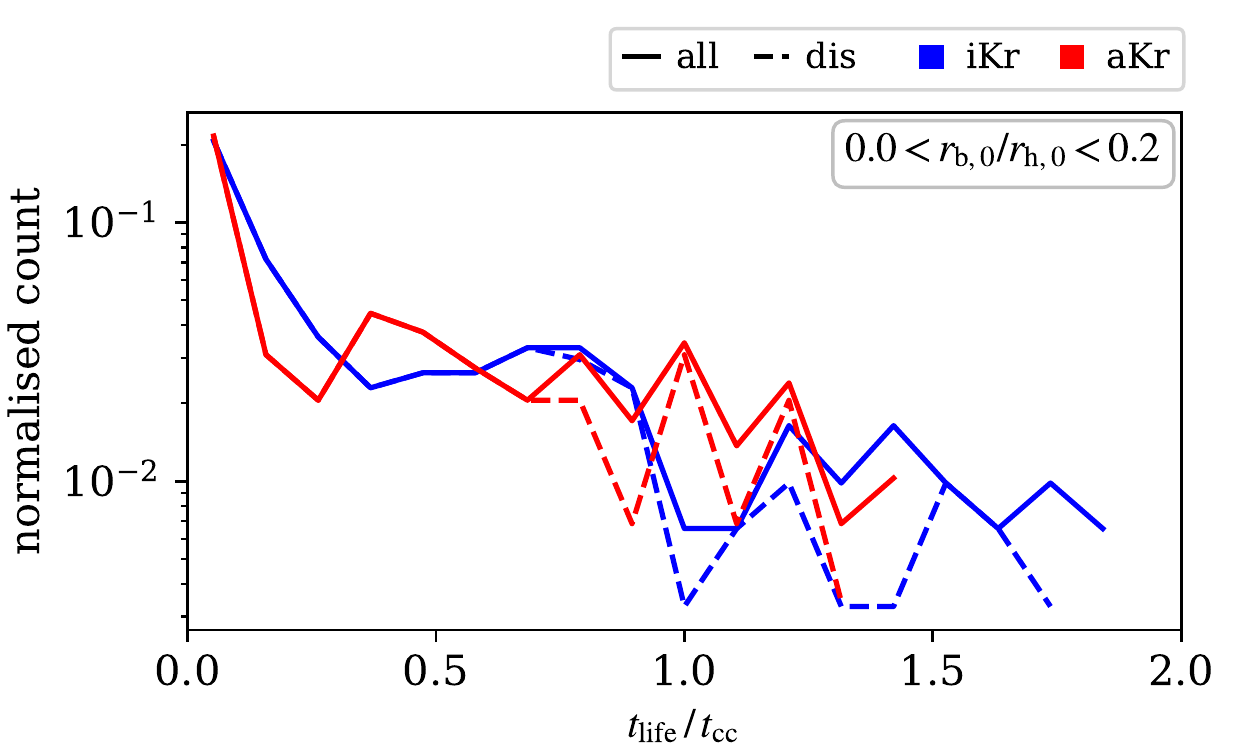}\\
	\caption{Lifetime of the primordial binary stars (allowing for the exchange of components one at a time, see text). Two cases are distinguished: total number of primordial binaries lost at a given time (solid lines, ``all''), and those that were disrupted or went through a double exchange between two model outputs (dashed lines, ``dis''). Hence the difference between the dashed and the solid lines is the number of primordial binaries that were ejected as bound pairs. All binaries are separated in three radial shells based on their initial position in the SC and the histograms are normalised by the total number of binary stars that were present in a given shell initially. Individual models are colour-coded.}
	\label{fig:lifetime}
\end{figure}

To further support this finding, we show the distribution of lifetimes of the primordial binaries in Fig.~\ref{fig:lifetime}, distinguished by their initial position in the cluster. The figure shows the distributions for all lost binaries and for those that were disrupted. The small difference between these curves indicates that the SCs mainly lose binaries due to disruptions and just a small fraction is escaping as bound pairs. It is only around core collapse when the number of escapers starts to rise due to an increasing number of dynamically induced ejections.

The lifetime of binaries which originated in the core or in the intermediate regions is independent of the model (see the bottom two panels of Fig.~\ref{fig:lifetime}).
This is to be expected since both systems are initially isotropic in the centre. There is, however, a significant difference in the lifetimes of primordial binaries that come from the outer regions (see the separation of the curves of both models in the top panel). The initially anisotropic cluster loses approximately twice the amount of halo binaries than the isotropic cluster at any given time.
In the anisotropic initial conditions a larger fraction of binaries from the outer regions are on radial/high-eccentricity orbits. Those systems periodically visit the central regions where they are subject to more frequent interactions, resulting in an enhanced rate in binary ionisation or ejection.

\subsection{Exchanges of components}

\begin{figure*}
	\includegraphics[height=18em,trim=0 0 67.6em 0,clip]{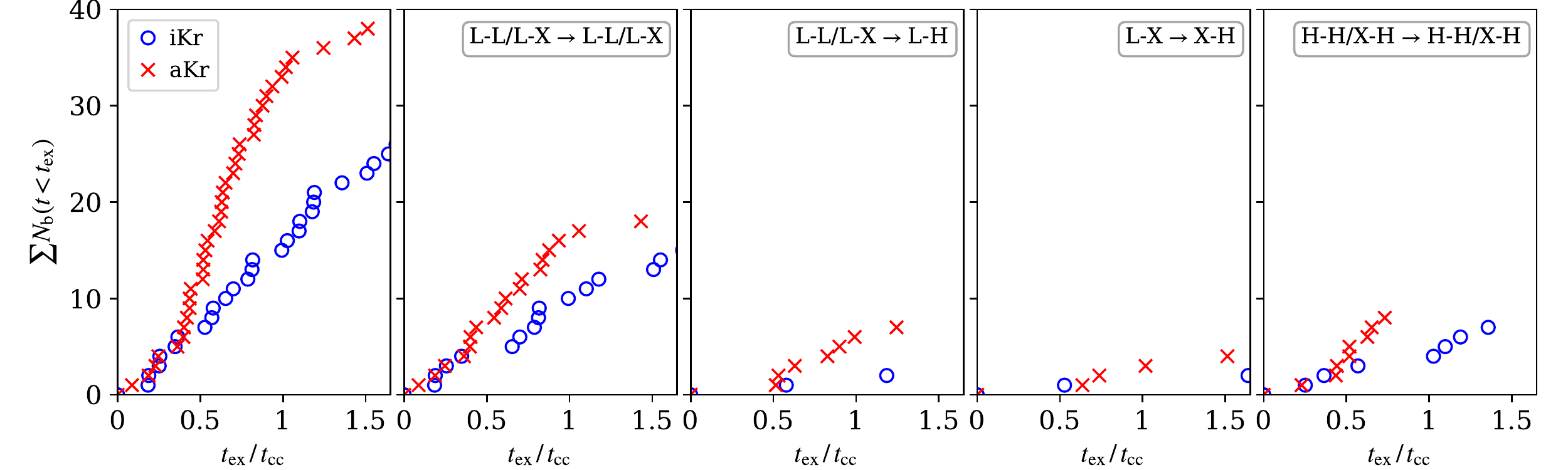}
	\hspace{1em}
	\includegraphics[height=18em,trim=22.7em 0 0 0,clip]{prim_life_before_1st_exch_outer_Kr}
	\caption{Time before the primordial binaries, which were initially in the outer regions ($2.0 < \rb[,0]/\rh[,0] < 10.0$), went through the first exchange of one of their components. Each point represents one binary system and the cumulative sum is plotted on the vertical axis. \textbf{Left panel:} All exchanges are plotted. \textbf{Four right panels:} The binaries are separated based on the mass groups in which their components belonged before and after the exchange (see text for the mass ranges definition). Binaries that would exchange both components and consequently end up disrupted are not included in this plot (see Fig.~\ref{fig:lifetime} instead).}
	\label{fig:prim_1st_Kr}

	\vspace{\floatsep}

	\includegraphics[height=18em,trim=0 0 33.5em 0,clip]{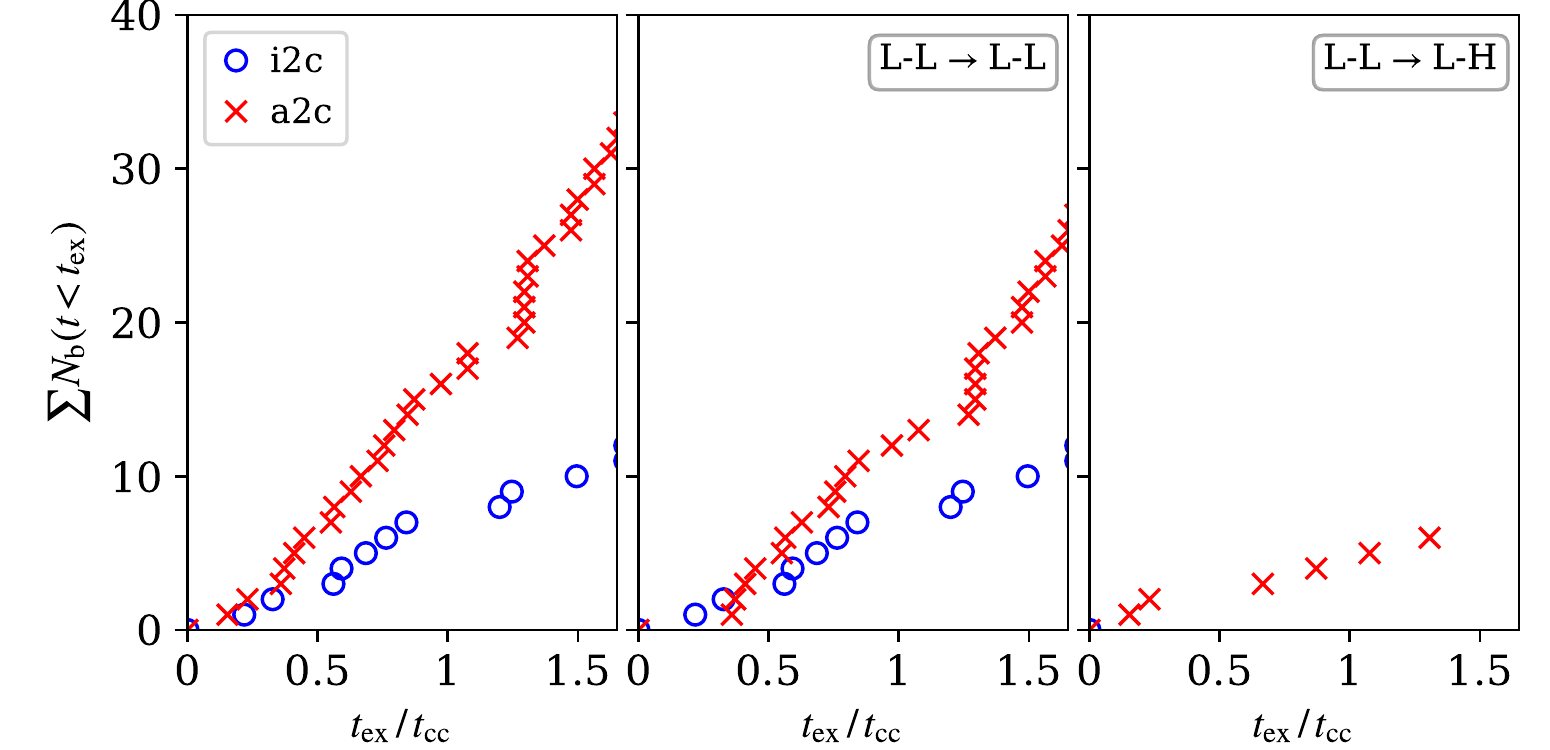}
	\hspace{1em}
	\includegraphics[height=18em,trim=22.7em 0 0 0,clip]{prim_life_before_1st_exch_outer_2c}
	\caption{Same as Fig.~\ref{fig:prim_1st_Kr} but for the two-component model.}
	\label{fig:prim_1st_2c}
\end{figure*} 
\begin{figure}
	\includegraphics[width=\colfigwidth,trim=0 17 0 0,clip]{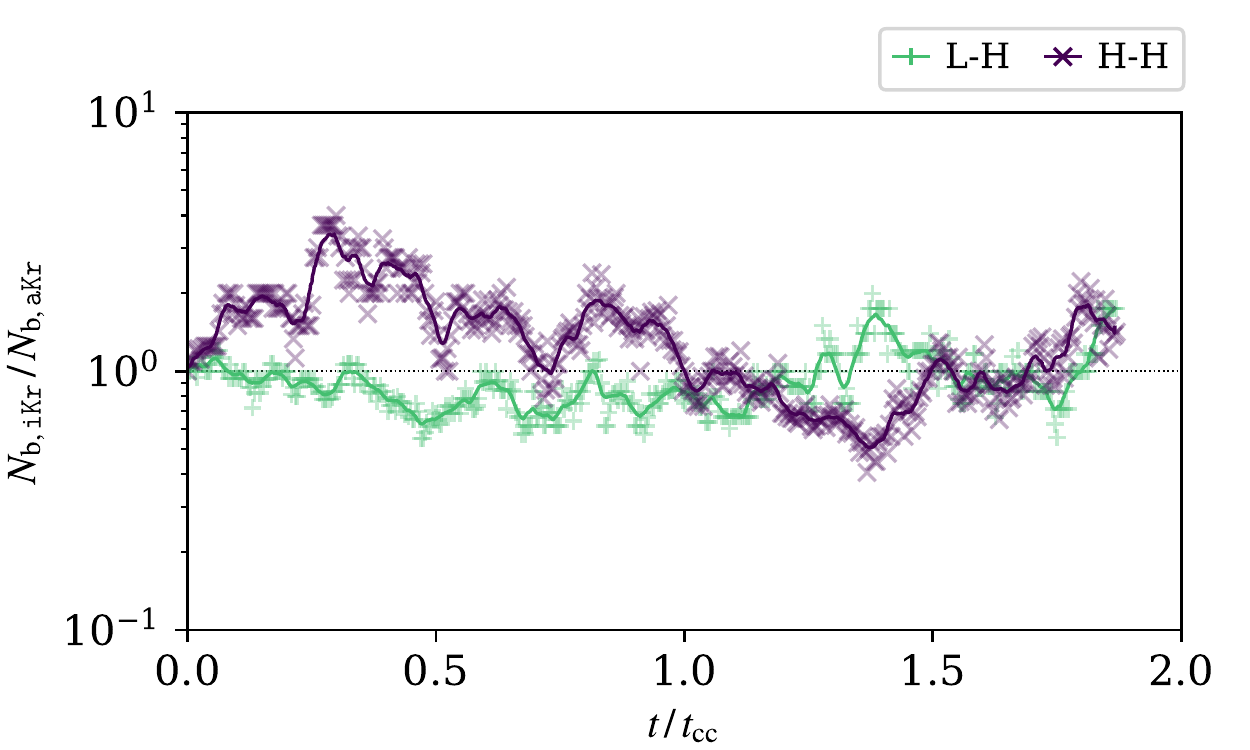}\\
	\includegraphics[width=\colfigwidth,trim=0 0 0 27,clip]{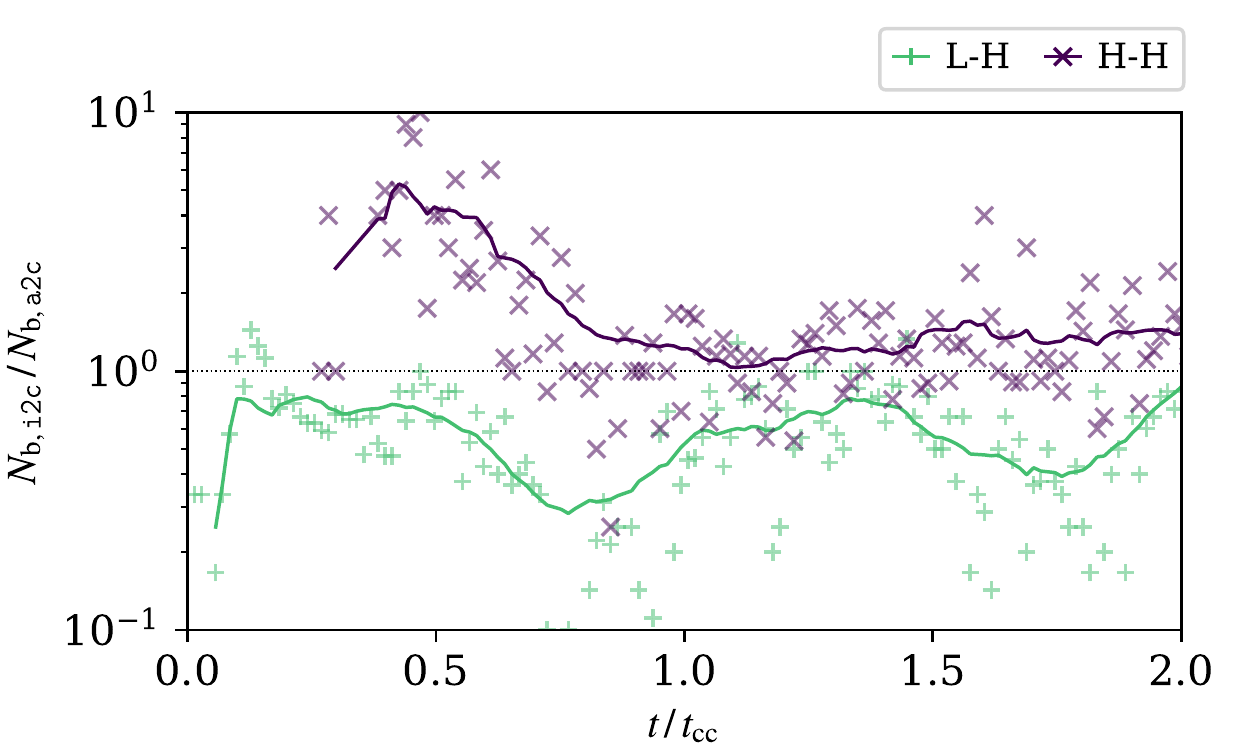}
	\caption{Ratio of the number of L-H (green pluses) and H-H (purple crosses) binaries in the isotropic vs anisotropic models, with Kroupa IMF (top) and two mass components (bottom). Time is scaled to the time of core collapse. The simple moving average is plotted by the thin line.}
	\label{fig:ratio_LH-HH_tcc}
\end{figure}

In this section we discuss the rate of encounters that lead to an exchange of binary star components. Alongside the \texttt{iKr} and \texttt{aKr} models with a continuous stellar mass spectrum, here we are also using two specifically designed two-component models -- \texttt{i2c} and \texttt{a2c}. These allow us to explore the rate of exchange events in a more simplified context and also carry out a preliminary study of the possible implications for the formation of binary systems including stellar remnants.

The distinction between the high-mass (``H'') and the low-mass stars (``L'') in the \texttt{i2c} and \texttt{a2c} models corresponds to their mass components (see Sect.~\ref{sec:methods}). In the \texttt{iKr} and \texttt{aKr} models, we keep the previously defined mass ranges (see Sect.~\ref{sec:segr}) -- i.e.\ \hbox{$0.9 \leq m_\mathrm{H}/\Msun \leq 1.0$}, and \hbox{$0.1 \leq m_\mathrm{L}/\Msun \leq 0.2$}) and also use an intermediate-mass group (``X'', with \hbox{$0.2 < m_\mathrm{X}/\Msun < 0.9$}).

We note that although the results from the previous section were only discussed for the models with a continuous IMF, we found qualitatively similar results for the two-component models as well. The only difference is in the much shorter evolutionary time scale of the two-component models, therefore, to make a comparison between the \texttt{Kr} and \texttt{2c} models, we scale the evolution to their times of core collapse.

In the left-hand panels of Figs.~\ref{fig:prim_1st_Kr} \&~\ref{fig:prim_1st_2c} we show the time before a primordial binary goes through the first non-disruptive exchange of one of its components. In both plots, we only include the events that occurred in the outer regions because, as we showed, those are most affected by the differences in the initial velocity distribution.\!\footnote{We note that the exchange rates in the inner regions show no differences between the isotropic and anisotropic models, therefore their plots are not included.}
The number of exchange events is growing more rapidly for the \texttt{aKr} and \texttt{a2c} models than their isotropic counterparts. As shown in the right-hand panels of both figures, this holds for all the possible exchange events, regardless of the masses of the components before and after the exchange. These results provide further evidence of the enhanced dynamical activity of binaries in the SCs with initial radially anisotropic velocity distribution.

As we further show in Fig.~\ref{fig:ratio_LH-HH_tcc} the anisotropic models typically have more L-H binaries than the isotropic ones, and conversely, the isotropic models have more H-H pairs at any given time. This holds until core collapse when the binary production becomes equivalent but also more noisy.

Both of these results are another manifestation of the effects of the anisotropic velocity distribution:
\begin{enumerate}
	\item	Due to the delayed mass segregation in the inner regions of the anisotropic models, there are fewer H stars in the core and, consequently, fewer opportunities for the formation of H-H binaries.
	\item	The greater number of binary stars on radial and highly elliptical orbits in the anisotropic models imply that these pairs can quickly reach the core where the inner H stars are segregating. Some of these in-falling binaries go through an interaction during which a high-mass star replaces one of its former L components.
	After such an encounter, the newly formed L-H binary may continue on an outward trajectory, leave the core, and move towards the outer regions. This mechanism not only depletes the core of the anisotropic cluster of high-mass stars, which makes it more challenging for the H-H binaries to form there, but it also removes the new L-H binaries from the dynamically active regions, enabling them to survive longer.
	\item	The isotropic cluster segregates its H stars towards the core on a shorter timescale, and there are few low-mass binaries in the core that can extract them from this region. Consequently, the number of L-H binaries is smaller, they cannot survive as long, and the production of H-H binaries can happen sooner.\!\footnote{Despite the initial kinematical differences the total number of H-H binaries is strictly limited by the SC population and stellar masses, which determine the depth of core collapse and the system's need to produce very hard binaries that can stop the central contraction \citep[cf.][]{tanikawa,fujii_pz,pavl_subr}. In all our models, the number of H-H binaries saturates between 10 and 15 after core collapse, with only minor fluctuations.}
\end{enumerate}

\section{Conclusions}
\label{sec:concl}

In this work, we investigated the role of the initial velocity anisotropy in the long-term dynamical evolution of star clusters, specifically focusing on two fundamental aspects: mass segregation and the dynamics of primordial binary stars.
We showed that both of them are influenced by the kinematical properties of the initial conditions and the underlying differences between the distributions of stellar orbits in the isotropic and anisotropic systems.
The collisional effects of the cluster are enhanced in the anisotropic system since a larger fraction of its outer stars are on radial/highly elongated orbits.
Our semi-analytical calculations and $N$-body simulations revealed that:
\vspace{-6pt}
\begin{enumerate}
	\item	the mass segregation time scale is longer in the inner regions and shorter in the outer regions of the anisotropic system when compared to the isotropic system;
	\item	the disruption rate of primordial binaries that were initially located in the outer regions is higher in the anisotropic clusters than in the isotropic ones;
	and
	\item	the rate of binary ejections and exchange encounters in which one of the binary primordial components is replaced by another interacting star is higher for outer primordial binaries in anisotropic clusters (no such effect was found for the inner binaries).
\end{enumerate}

Our results have a broader impact on the general understanding of the dynamical evolution of stellar systems.
In particular, the differences in the degree of mass segregation affect the evolution towards core collapse and the extent of the internal radial variation of the stellar mass distribution.
The enhanced binary activity in the anisotropic systems has implications for the properties of the surviving binaries, and for the present-day binary fraction. This also includes exotic binaries (e.g.\ X-ray binaries, blue stragglers, close white dwarf--main sequence binaries, millisecond pulsars or gravitational wave sources) resulting from dynamical interactions and exchanges in the dense environment of globular clusters.
Future efforts aimed at building more realistic models will require the inclusion of the effects associated with stellar evolution and a comprehensive exploration of different initial properties (e.g.~different King concentrations, filling factors, orbits in the host galaxy).
We will further explore these issues in the follow-up studies.

\section*{Acknowledgements}

We thank the Stellar/Galactic group at the Dept. of Astronomy for valuable discussions. VP is also grateful to Steven Shore for enlightening conversations.
This research was supported in part by Lilly Endowment, Inc., through its support for the Indiana University Pervasive Technology Institute,
in part by Shared University Research grants from IBM, Inc., to Indiana University,
and in part by the Indiana METACyt Initiative.
VP also acknowledges the use of the high-performance storage within his project \textit{``Dynamical evolution of star clusters with anisotropic velocity distributions''} at Indiana University.
The \texttt{Python} programming language with \texttt{NumPy} \citep{numpy} and \texttt{Matplotlib} \citep{matplotlib} contributed to this project.
This research has made use of NASA's Astrophysics Data System Bibliographic Services.

%

\section*{Data availability statement}

The data presented in this article may be shared on reasonable request to the corresponding author.



\bibliographystyle{mnras}
\bibliography{bibliography} 

\bsp	
\label{lastpage}
\end{document}